\documentclass[fleqn,12pt,twoside]{article}

\usepackage{espcrc1}

\usepackage{graphicx}

\usepackage{floatflt}

\newcommand{\snn}{\sqrt{s_{NN}}}
\newcommand{\seff}{\s_{\rm eff}}
\newcommand{\s}{\sqrt{s}}
\newcommand{\AAA}{A+A}
\newcommand{\pp}{p+p}

\newcommand{\epem}{e^+e^-}

\newcommand{\nch}{N_{ch}}
\newcommand{\np}{N_{part}}

\newcommand{\ntot}{\langle\nch\rangle}

\newcommand{\nc}{N_{coll}}

\newcommand{\halfnp}{\langle\np/2\rangle}
\newcommand{\etap}{\eta^{\prime}}

\newcommand{\etaone}{|\eta| < 1}
\newcommand{\dndeta}{d\nch/d\eta}
\newcommand{\dndetap}{d\nch/d\etap}

\newcommand{\dndetaone}{\dndeta|_{\etaone}}
\newcommand{\dndetanp}{\dndeta / \halfnp}
\newcommand{\dndetapnp}{\dndetap / \halfnp}

\newcommand{\dndetaonenp}{\dndetaone / \halfnp}
\newcommand{\ratio}{\ntot/\halfnp}

\newcommand{\yb}{y_{\rm beam}}
\newcommand{\mpt}{\langle p_T \rangle}

\title{Bulk Dynamics in Heavy Ion Collisions}

\author{Peter A. Steinberg 
\thanks{Current address: Chemistry Department, Brookhaven National Laboratory, Upton, NY 11973}
}
\begin{document}

\maketitle

\begin{abstract}
The features of heavy ion collisions that suggest the relevance of
collective dynamics, as opposed to mere superpositions of nucleon-nucleon
or even parton-parton collisions, are reviewed.  The surprise of these
studies is that bulk observables are far simpler than typical dynamical
models of nucleus-nucleus collisions would imply.
%
%
%
%
These features are shown to have a natural interpretation in terms
of statistical-hydrodynamical models.
%
%
%
%
The relevance of hydrodynamics to heavy ion collisions, coupled
with the various similarities of the heavy ion data with that
of more elementary collisions, raises very basic questions
about its relevance to smaller systems.  
\end{abstract}

\section{Introduction}

Recently, there has been dramatic progress in the understanding
of the dynamics of heavy ion collisions.  This is due to the
availability of a large, high-quality data set spanning an
enormous range in energy, rapidity and event centrality.
With the turn-on of the RHIC facility in 2000, we now have 
information about the basic features of these collisions up to
$\snn=200$ GeV, where $\snn$ is the nucleon-nucleon center-of-mass
energy.  Although most of the analyzed data is from mid-rapidity
(90 degrees in the center-of-mass system), all of the accelerator facilities
have experiments dedicated to acquiring data over the full rapidity
range (e.g. E895 at the AGS, NA49 at the SPS, and BRAHMS and PHOBOS at RHIC).  
Finally, the various heavy ion experiments are starting to
converge on comparable centrality measures, making comparisons
between the data sets as a function of the number of participating
nucleons feasible.

During the summer of 2004, the four RHIC experiments (BRAHMS, 
PHENIX, PHOBOS, and STAR) produced draft ``white papers'', summarizing 
their most important results and relevant interpretations~\cite{Arsene:2004fa,Back:2004je,Adcox:2004mh,STAR-wp}.
The result of these discussions indicates an ongoing
paradigm shift both in the
interpretation of the lattice results and the experimental data.
The 20-25\% deviation of the lattice energy density from the
Stefan-Boltzmann Limit is no longer interpreted as approaching
the weakly-interacting Quark Gluon Plasma.  Rather, this deviation
is presently understood (using N=4 SUSY QCD) as precisely the
signature of a strongly-interacting plasma~\cite{Gubser:1996de,Shuryak:2004rh}.
Thus, rather than expecting a weakly-coupled gas of quarks and
gluons to be liberated in high energy heavy ion collisions, 
one might expect to create a strongly-interacting state instead,
with completely different properties than previously expected.
More importantly, one would
not expect systems with these properties to be generated by the
two body scatterings of asymptotic hadrons.  The energy density
achieved in these collisions imply particle densities far too
high for individual particle states to remain distinct.

\begin{figure}[t]
\begin{minipage}{60mm}
\includegraphics[width=60mm]{fig2_qm02.epsi}
\caption{$\dndetaonenp$ compared with hadronic and pQCD-based
models.
\label{fig2_qm02}}
\end{minipage}
\hspace{\fill}
\begin{minipage}{90mm}
\includegraphics[width=90mm]{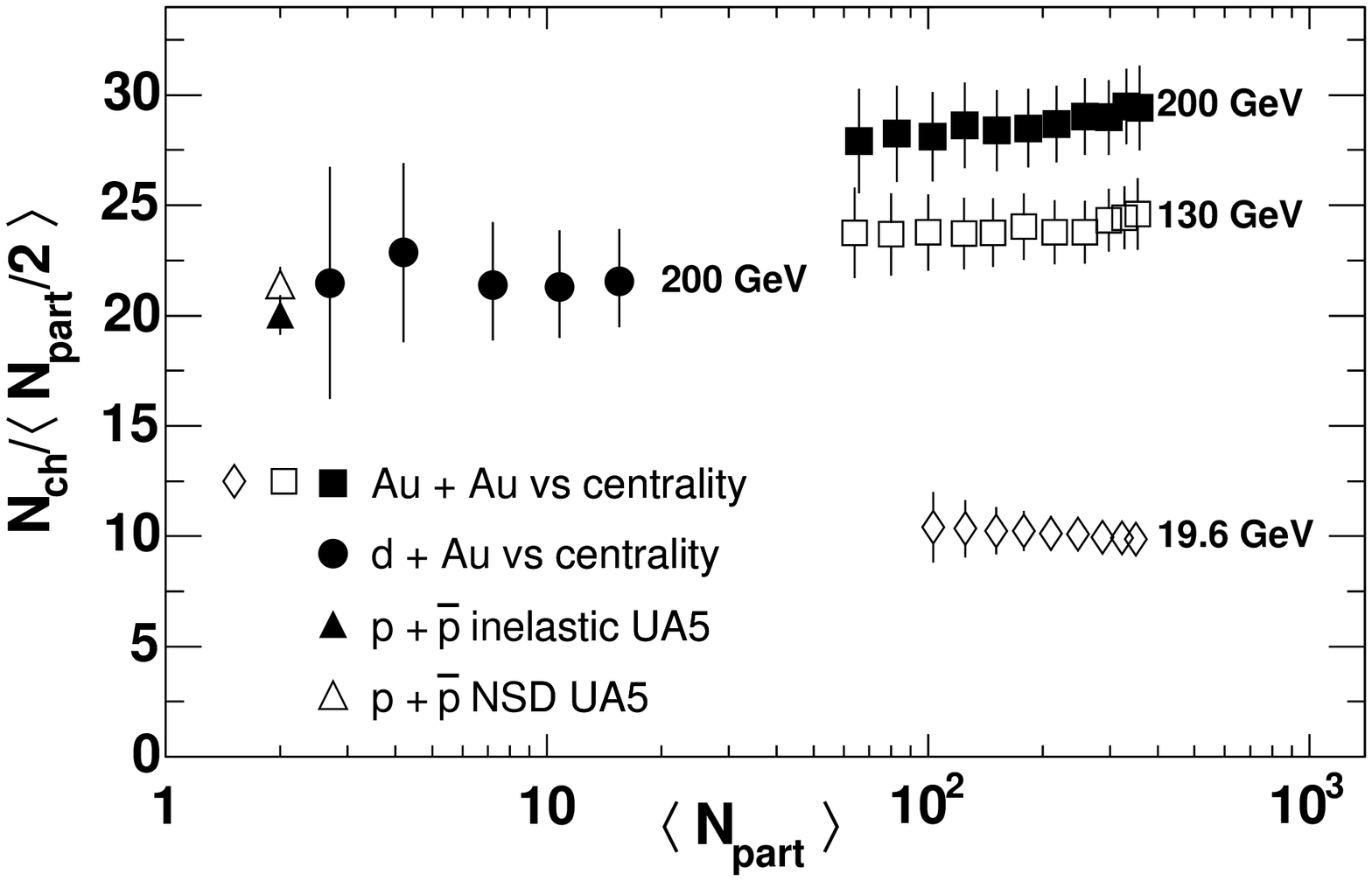}
\caption{
Total number of charged particles per participant pair as a
function of collision centrality for three RHIC energies. 
d+Au data from $\snn=200$ GeV is also shown.
\label{WP12_Ntot_Npart_AuAu_19130200_dAu_pp}
}
\end{minipage}
\end{figure}

If this is the case, then one should expect the system created
in heavy ion collisions to show collective behavior, characteristic
of a system with zero mean free path, i.e. the hydrodynamic limit.
This treatment of the system stands in stark contrast to the usual dynamical
transport approach to heavy ion collisions.  In this sort of approach,
one expects multiple, independent stages, characterized by 
different dynamical mechanisms
(shadowed parton distributions, parton production and reinteractions,
quark recombination and chemical freezeout, fragmentation functions,
hadron rescattering, thermal freezeout and hadronic decays).

The independent combination of these various sources is
encoded in models such as HIJING.  There are also models that
avoid the partonic stage altogether and proceed directly to the hadronic
dynamics, such as RQMD.  Interestingly, all of these models
have difficulty reproducing the energy dependence of the particle
density at mid-rapidity ($\dndetaonenp$) 
as shown in Fig.~\ref{fig2_qm02}~\cite{ToporPop:2002gf}, which compares predictions
for Au+Au and $\pp$ from HIJING and RQMD, and compares them with data.
While the models are tuned appropriately on $\pp$, both miss the
Au+Au data systematically as a function of beam energy.
Variables such as this integrate over
all of the possible dynamical processes outlined above, thus
showing the utility of bulk observables to be sensitive to 
both hard and soft processes, and thus early and late stages.
Of course, a single observable at a particular energy and centrality
is of little use.  Rather, it is the combined systematics of
energy, centrality, and rapidity which provide
a handle on the most important dynamical contributions.

The surprising conclusion, partially outlined in this proceedings,
is that the systematics of bulk observables are
far simpler than one might expect given the various mechanisms that
could contribute to the final state.  
However, it seems that one must consider the features of particle
production over the full phase space (and not just concentrate on
particular kinematic regions) to isolate the simple
organizing features.

\section{Features of charged-particle multiplicities in $\AAA$ collisions}

\begin{figure}[t]
\begin{minipage}{70mm}
\includegraphics[width=70mm]{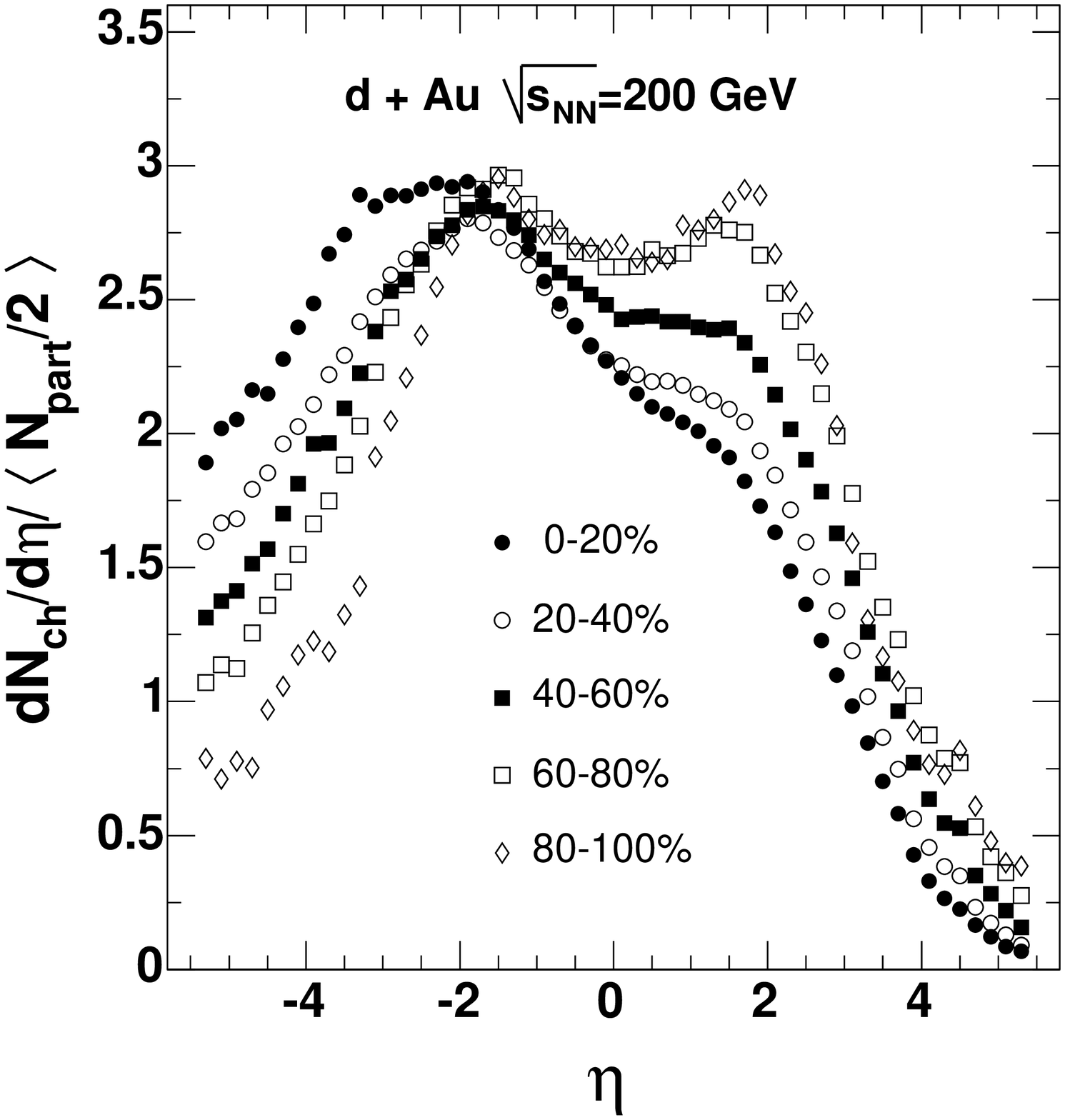}
\caption{
$\dndetanp$ in d+Au collisions at $\snn=200$ GeV as a function
of collision centrality.
\label{WP16_dNdetaNpart_eta_dAu_5cent}
}
\end{minipage}
\hspace{\fill}
\begin{minipage}{80mm}
\includegraphics[width=80mm]{WP14_dNdetaNpart_eta_AuAu_19_200.eps}
\caption{
$dN/d\eta/\halfnp$ for peripheral 
and central events at $\snn=19.6$ and $200$ GeV.
\label{WP14_dNdetaNpart_eta_AuAu_19_200}
}
\end{minipage}
\end{figure}

The most striking feature of bulk particle production in heavy
ion collisions is the approximate linearity of the total number
of charged particles ($\nch$) 
with the number of participating nucleons ($\np$).
Results on $p+A$ collisions in the 
1970's and 80's~\cite{Elias:1978ft}, found that
$\nch$ scaled linearly as $\np \times \nch^{pp}$.
Recent PHOBOS data on d+Au collisions, shown in Fig. ~\ref{WP12_Ntot_Npart_AuAu_19130200_dAu_pp}~\cite{Back:2004mr} 
shows that this phenomenon
persists at RHIC energies of $\snn=$200 GeV.
This simple behavior occurs despite non trivial changes to the
particle density over the whole pseudorapidity range,
shown in Fig.~\ref{WP16_dNdetaNpart_eta_dAu_5cent}.
What was not expected was that the same scaling
would be present in Au+Au collisions from $\snn=$ 20 to 200 GeV.
This is a non-trivial result, in that the variation of $\dndeta$
with centrality, shown in Figs.~\ref{WP14_dNdetaNpart_eta_AuAu_19_200},
still maintains an overall constancy of $\ratio$.
It should also be pointed out that while the d+Au data connects
simply to the $\pp$ data, the $\AAA$ data does not seem to 
extrapolate smoothly to the $\pp$ limit.

\begin{figure}[t]
\begin{center}
\begin{minipage}{80mm}
\includegraphics[width=80mm]{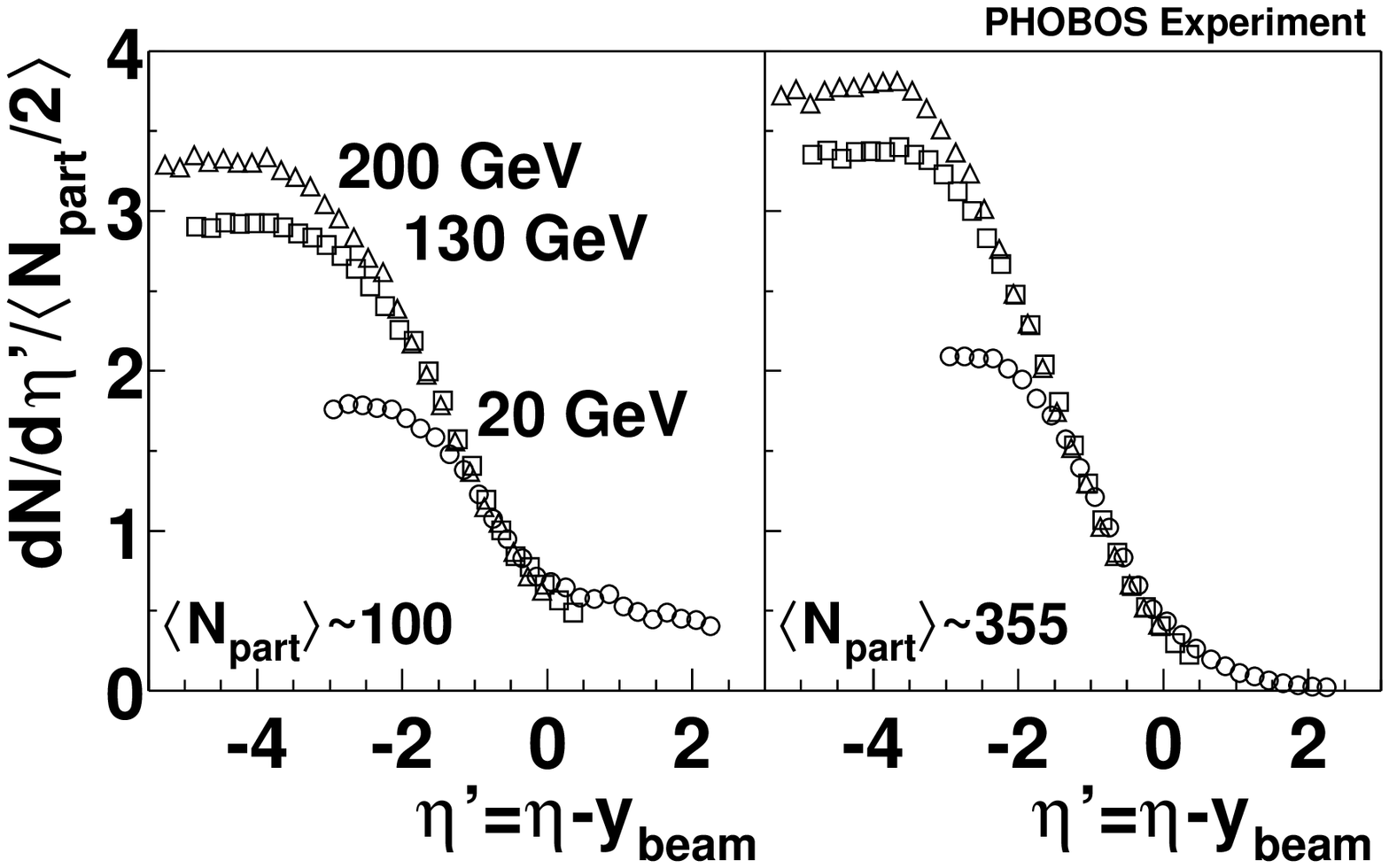}
\caption{$dN/d\etap/\halfnp$ for peripheral 
and central events at three RHIC energies.
\label{two_view}
}
\end{minipage}
\hspace{\fill}
\begin{minipage}{70mm}
\includegraphics[width=70mm]{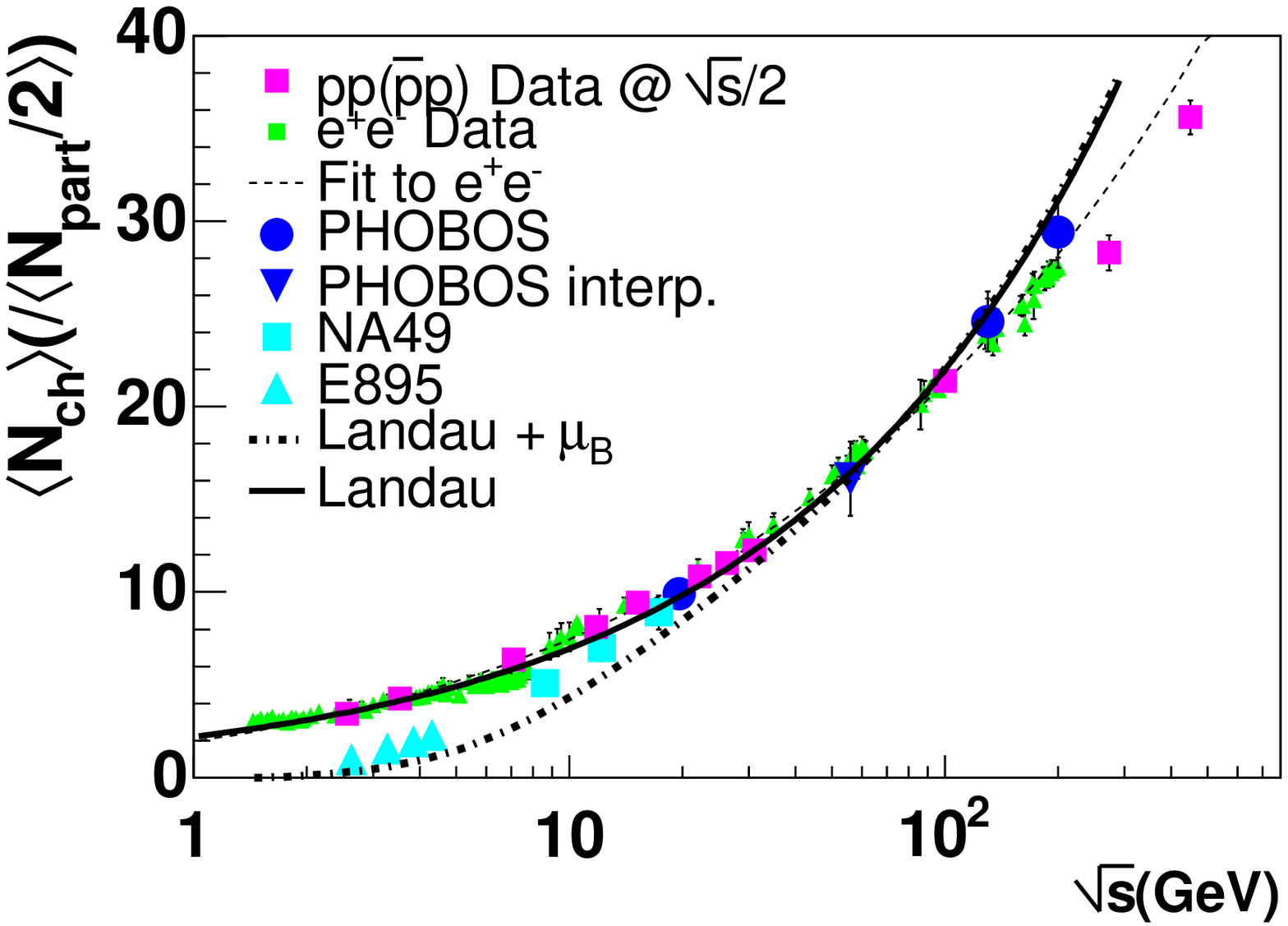}
\caption{
$\nch$ compared for $\AAA$, $\epem$ and $\pp$ collisions (both
at $\s$ and $\s/2$.  The data is compared to formulas described in the text.
\label{total_AA_ee_pps_landau_mub}
}
\end{minipage}
\end{center}
\end{figure}

Another striking global feature is ``limiting fragmentation'', the
energy independence of the particle yields with energy when
considering a system with a fixed collision geometry in a frame where
one of the projectiles is at rest~\cite{Back:2002wb} This feature was
expected to be found in very limited regions of pseudorapidity,
characteristic of the fragmentation of each
projectile~\cite{Benecke:sh}.  Instead, limiting fragmentation is
observed over a relatively wide range in pseudorapidity, as shown in
Fig.~\ref{two_view}.  Moreover, one finds that the shape of the
``limiting curve'' depends only on the impact parameter, and once this
is determined, data from different energies lies on the same curve.
Since the data peels off of the limiting curve at approximately the
same distance from mid-rapidity, this phenomenon also seems to
strongly constrain the mid-rapidity yields as well.  Thus, the energy
dependence of particle yields seems to be controlled by a global
constraint that is obeyed at all available beam energies.

The third striking feature of bulk particle production
is the apparent universality of the total multiplicity for
$\AAA$, $\pp$ and $\epem$ reactions at high energy, shown in
Fig.~\ref{total_AA_ee_pps_landau_mub}~\cite{Back:2003xk,Steinberg:2004vy}.  
Of course,
this is only observed if one accounts for the ``leading particle effect''
in $\pp$ collisions by working with an ``effective energy''
$\seff=\s/2$.
The agreement of $\AAA$ and $\epem$ without rescaling suggests on
the contrary that there is no similar leading particle effect
in heavy ion collisions.  This is presumably connected to the fact that
a typical participant is struck multiple times in $\AAA$ (unlike $\pp$
and $p+A$ collisions).
The deviation of $\AAA$ and $\epem$ at low energies can be understood
as a consequence of the presence of a substantial baryochemical potential~\cite{Steinberg:2004vy},
which tends to suppress the overall entropy (via the relationship
$S=(E-pV-\mu_B N_B)/T$.

In conclusion, there seem to be three essential features of particle
multiplicities in $\AAA$ collisions: 1) $\np$-scaling of the total multiplicity,
2) A universal value of $\ratio$ in $\AAA$, $\pp$ and $\epem$ reactions,
and 3) ``limiting fragmentation'', which seems to constrain the global
angular distributions and thus even the mid-rapidity density.
Such simple behavior would not naturally be expected to arise out of
a dynamical approach, where each stage is in-principle distinct from 
the others.  For instance, it would seem fortuitous for nuclear shadowing,
energy loss, and fragmentation functions to conspire to arrive
at $\np$-scaling from processes where semi-hard physics is controlled
by the number of binary collisions, $\nc$.  It seems that any relevant
physics scenario (such as the Color Glass Condensate described by
L. McLerran~\cite{McLerran:INPC})
must capture these essential features.

\section{Hydrodynamic Descriptions of High Energy Collisions}

As discussed in the introduction, one of the major surprises
from the RHIC data has been the relevance of relativistic 
hydrodynamics in the overall understanding of the experimental
data.  In some sense, this should not be so surprising.  It is
well-known that any system in which the mean free path approaches
zero should show hydrodynamic behavior, even an apparently
disparate system such as cold $^6$Li atoms~\cite{ohara}.
This system can be initialized with a strongly asymmetric geometry
and the subsequent pressure gradients lead to spatial deformations
that have been successfully modeled with hydrodynamics.

\begin{figure}[t]
\begin{minipage}{80mm}
\includegraphics[width=80mm]{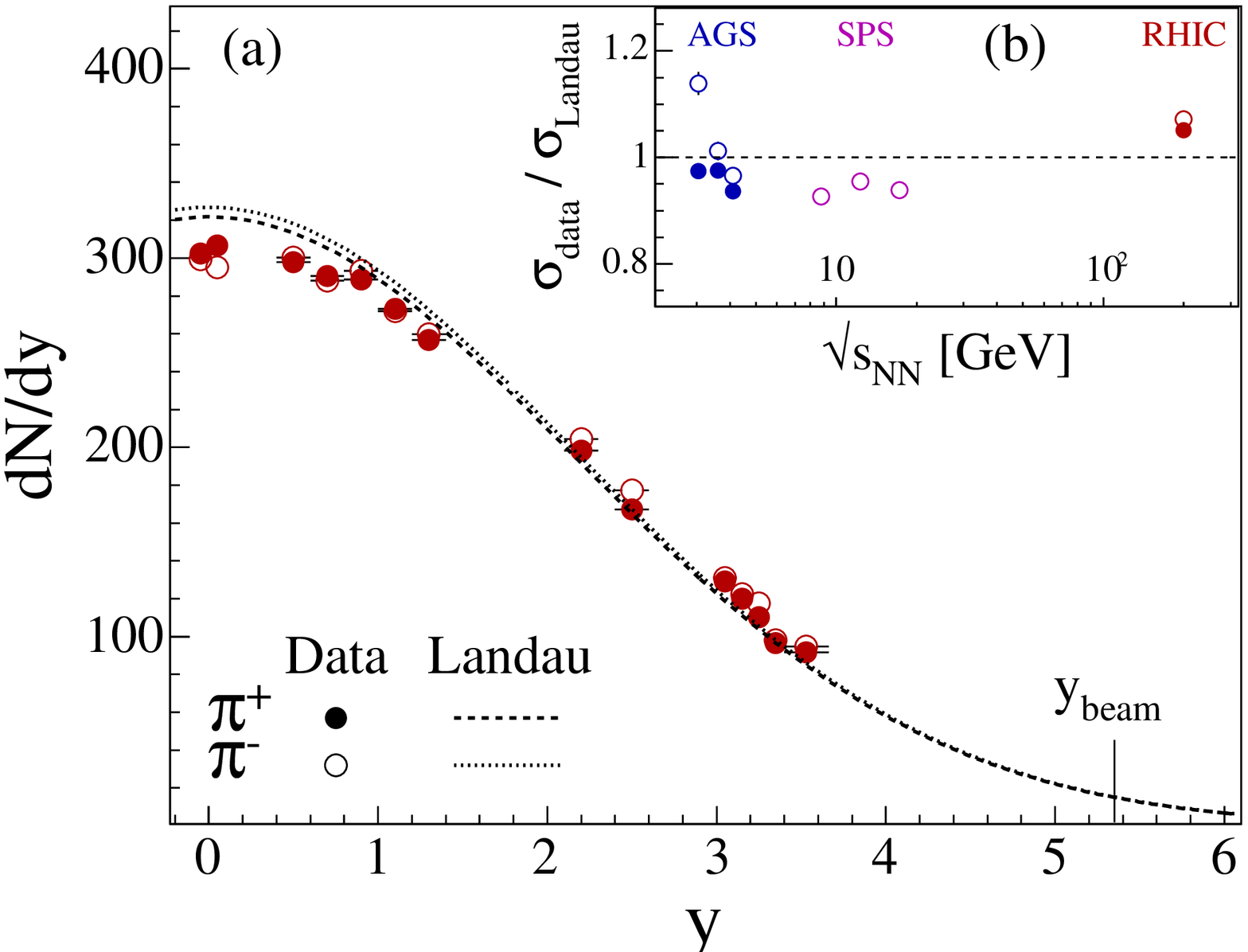}
\caption{
BRAHMS data $dN/dy$ for charged pions at 200 GeV, fit to a Gaussian.
The inset shows the width divided by the Landau expectation as
a function of $\snn$.
\label{landau}
}
\end{minipage}
\hspace{\fill}
\begin{minipage}{70mm}
\includegraphics[width=70mm]{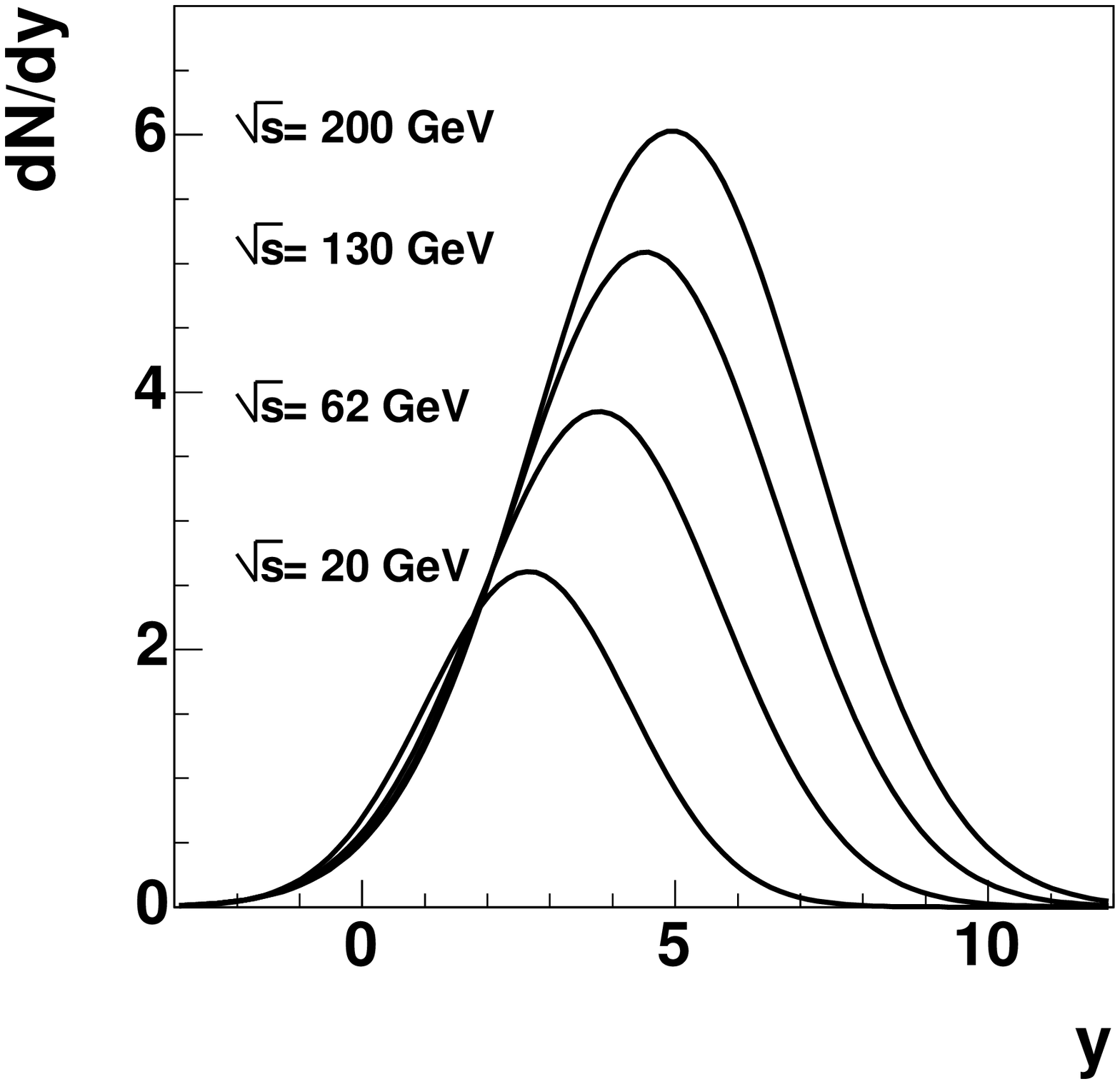}
\caption{
The Landau prediction for the multiplicity and rapidity distributions,
observed in the limiting fragmentation reference frame.
\label{landau_limfrag_2}
}
\end{minipage}
\end{figure}

The relevant geometry in the early stages of a heavy ion collision
(or even a $\pp$ collision) is one characterized by the nuclear
radius in the transverse direction, and a large contraction
in the longitudinal direction.  Already one sees that the relevant
time scales for longitudinal and transverse dynamics are very
different, of order $\tau_T \sim R/c_s$ (approximately several
fm/c) in the transverse direction
and $\tau_L \sim mR/(c_s \s)$ ($\ll 1$ fm/c) in the longitudinal direction.

\subsection{Landau Hydrodynamics}
Already in the early 1950's, 
Fermi and Landau considered a system with the geometry
just described, replacing the potential complexities of a high energy
nuclear collision with a slab of area $\pi R^2_A$ and length
$\Delta = R_Am/\s$ (and thus volume $V\propto R^3/\snn$) 
in which all of the energy of the incoming
projectiles is assumed to thermalize. This leads to an 
initial energy density of $\epsilon \propto s_{NN} \sim 4$ TeV/fm$^3$ at RHIC
energies, a value typically seen as unphysically large.
And yet, when one uses the massless blackbody equation of state 
$p=\epsilon/3$ and converts this to a relation for the entropy density, 
$\sigma \propto \epsilon^{3/4}$,
the total entropy in the collision volume is 
$S=\sigma V \propto s^{3/4}/\s = s^{1/4}$, which is itself
proportional to the 
total multiplicity~\cite{Fermi:1950jd,Landau:gs,Belenkij:cd}.
It may not be immediately clear from these definitions, but in a
nuclear collision, $S \propto V \propto \np$, leading naturally
to the $\np$-scaling of total multiplicity.  
Landau extended this statistical model by
using the equations of relativistic hydrodynamics~\cite{Landau:gs,Belenkij:cd}.
It turned out that the strong compression
along one axis leads to highly anisotropic distributions, with 
Gaussian rapidity distributions of width $\sigma = \ln\sqrt{s/m^2}$.
It is the application of the Fermi-Landau initial conditions to
the generally-accepted formalism of 3D relativistic hydrodynamics
that is known as the ``Landau hydrodynamical model'', as
advanced by Cooper et al.~\cite{Cooper:1974ak},
Carruthers et al.~\cite{Carruthers:dw}
and Shuryak~\cite{Zhirov:qu}, among others.
What is striking is just how much existing data is broadly consistent
with Landau's original predictions from the 1950's. 
As shown in Fig.~\ref{total_AA_ee_pps_landau_mub}, the Landau 
multiplicity formula, tuned on the low-energy data 
$\nch = 2.2s^{1/4}$ does a reasonable job on describing the trend
of $\epem$ over a wide range of $\s$, and thus the $\pp$ data
at $\s/2$ and the $\AAA$ data above $\snn=20$ GeV~\cite{Steinberg:2004vy}.
Of course it should not be overlooked that pQCD~\cite{Mueller:cq} 
can describe
the $\epem$ data just as well, if not better, than the Landau
formula.  However, it is not clear why the two approaches agree
to better than 10\% over the range for which data exists, as shown
in Fig.~\ref{total_AA_ee_pps_landau_mub}.

The BRAHMS rapidity distributions of charged pions are distinctly 
Gaussian (shown in Fig.~\ref{landau})
and have a width which deviates less than $10\%$ from
Landau's parameter-free predictions (as seen in the inset of 
Fig.~\ref{landau})~\cite{Bearden:2004yx}.
It is less well-known that Landau's formulas ($\nch \propto s^{1/4}$
and $\sigma = \ln\sqrt{s/m^2}$) actually {\it predict} limiting
fragmentation when plotted as a function of $y^{\prime}=y-\yb$,
as shown in Fig.~\ref{landau_limfrag_2}~\cite{Steinberg:2004vy}.
It appears that Landau's physical picture is already consistent with
the essential features of particle multiplicities in heavy ion
collisions.
Moreover, in contrast to models that explain the rapidity distributions
as the consequence of scatterings between the partons in the initial-state
nucleon and nuclear wave functions~\cite{Kharzeev:2001gp}, the Landau
model starts with a static initial state and rapidly generates the
rapidity distribution by means of hydrodynamics.  Thus, $\dndeta$
is not a static ``initial-state'' effect, but rather the result of
a dynamical process in the very first stages of the collision.

\subsection{Bjorken Hydrodynamics}

\begin{figure}[t]
\begin{minipage}{75mm}
\includegraphics[width=75mm]{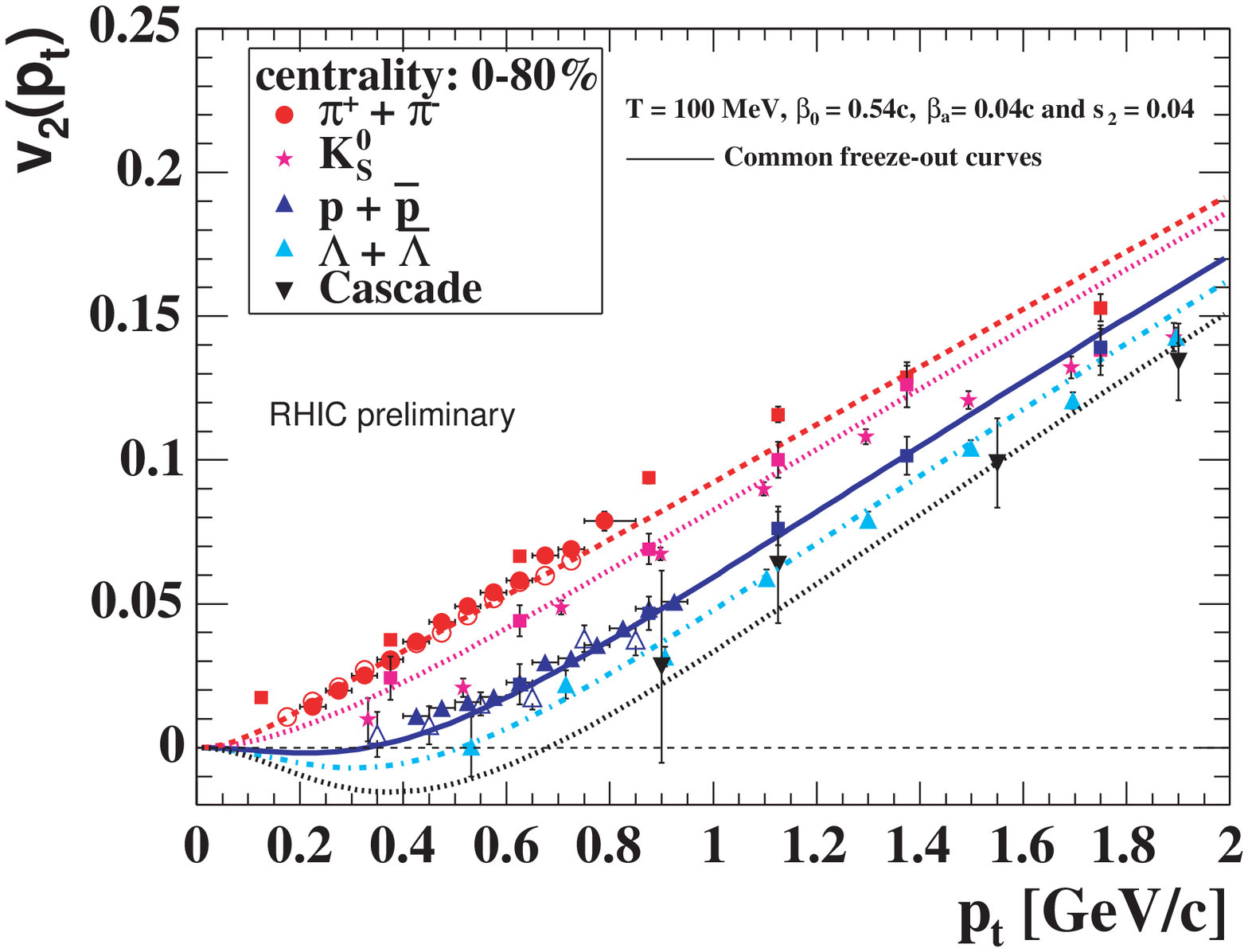}
\caption{
$v_2$ as a function of $p_T$ for various particle species,
compared to a hydrodynamics-inspired fit.
\label{raimond_prev}}
\end{minipage}
\hspace{\fill}
\begin{minipage}{75mm}
\includegraphics[width=75mm]{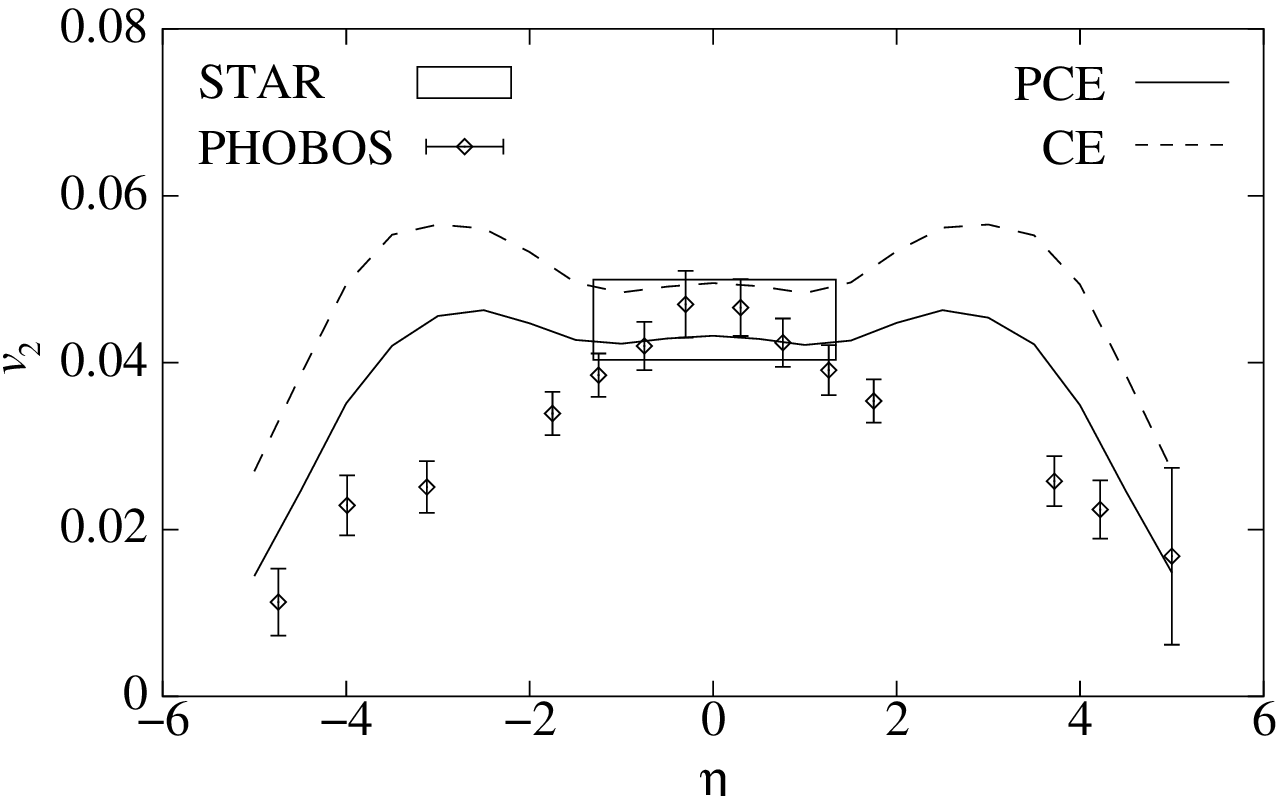}
\caption{
$v_2$ as a function of $\eta$ compared with 3D hydro
calculations.
\label{v2eta}
}
\end{minipage}
\end{figure}

Although Landau hydrodynamics appears to be relevant for the physics
in the very early stages ($\tau \ll 1 fm/c$), it is still a model
only understood semi-analytically with some particularly drastic
approximations.  By contrast, hydrodynamic calculations which
assume boost-invariance in the initial conditions have been 
used for quantitative comparisons with experimental observables
that are sensitive to early-time pressure gradients~\cite{Kolb:2003dz}.
These models are based on the pioneering work of Bjorken~\cite{Bjorken:1982qr}, who postulated the imposition of boost-invariance~\cite{Feynman:ej}
as a guiding principle for high energy reactions.
Of course,
since the calculations are initialized at time scales on the order of 1 fm/c,
they are unable to calculate the initial-state entropy. However,
given this single piece of experimental data, and an assumed equation of
state (usually a hybrid of the Landau EOS and a hadronic EOS, with a mixed
phase), they are able to successfully
calculate the effects of transverse pressure on particle spectra
(radial flow)
the mapping from the initial-state geometry to anisotropies
in the final-state transverse momentum distributions (elliptic flow)~\cite{Kolb:2003dz}.

Several clear signatures of radial flow are present in the
experimental data even without recourse to particular models.
At low transverse momentum ($p_T < m$) it is observed that
the particle spectra harden with increasing particle mass.
This is characteristic of a collective flow velocity field that 
gives heavier particles a larger momentum kick
($p=\gamma\beta m$).  This immediately breaks the $m_T$-scaling seen in
$\pp$ data, which is often thought to result from emission from a 
thermalized source.  Models which include a more detailed handling
of chemical freezeout are able to reproduce these trends\cite{Kolb:2002ve}.

Elliptic flow has now been comprehensively described by calculations
using boost-invariant hydrodynamics, both as a function of 
centrality and as a function of $p_T$ and particle mass for a fixed
centrality.  An example is shown in Fig.~\ref{raimond_prev}, which
shows the characteristic ``fine structure'', or mass-splitting of the
asymmetry parameter $v_2$ as a function of the particle's transverse
momentum~\cite{Snellings:2004ep}.
These are non-trivial relationships between various species that are not 
typically described well in the dynamical approach described above.
Nor can typical parton transport models reproduce the magnitude of
the asymmetries~\cite{Molnar:2004zj}.
However, given the obvious discrepancy between the assumption of
boost-invariance used in these calculations and the manifestly boost
non-invariant particle distributions shown by the BRAHMS data, 
it is not surprising to find that it is difficult to reproduce 
the dependence of $v_2$ on $\eta$ as measured by PHOBOS, 
shown in Fig.~\ref{v2eta}~\cite{Hirano:2001eu}.
This is true even for truly 3D hydrodynamic calculations, showing
the need to understand the initial conditions in some detail.
The current state of the art calculations~\cite{Hirano:2004rs}
use gluon structure functions calculated in the CGC framework,
which explicitly avoids the longitudinal dynamics of the Landau approach.

In conclusion, we have seen that the hydro approach appears to 
warranted by a wide range of data, although no existing model
or code can describe every detail correctly.  This is especially true
when considering longitudinal dynamics, which has not yet been fully 
incorporated into models that describe many features of
the transverse dynamics.  

\section{Relation to Elementary Systems}

Typically, the success of hydrodynamic models in heavy ion collisions
is attributed to their being larger than hadronic length scales,
which are thought to control the scattering processes by which the
system equilibrates.
And yet, it has already been shown that certain features of elementary 
collisions, such as the total entropy density, 
are manifestly similar to heavy ion collisions.  
Thus, it is not ruled out {\it a priori} to
seriously consider the relevance of hydrodynamics to smaller systems, 
such as $\pp$ and $\epem$.

\begin{figure}[t]
\begin{center}
\includegraphics[width=85mm]{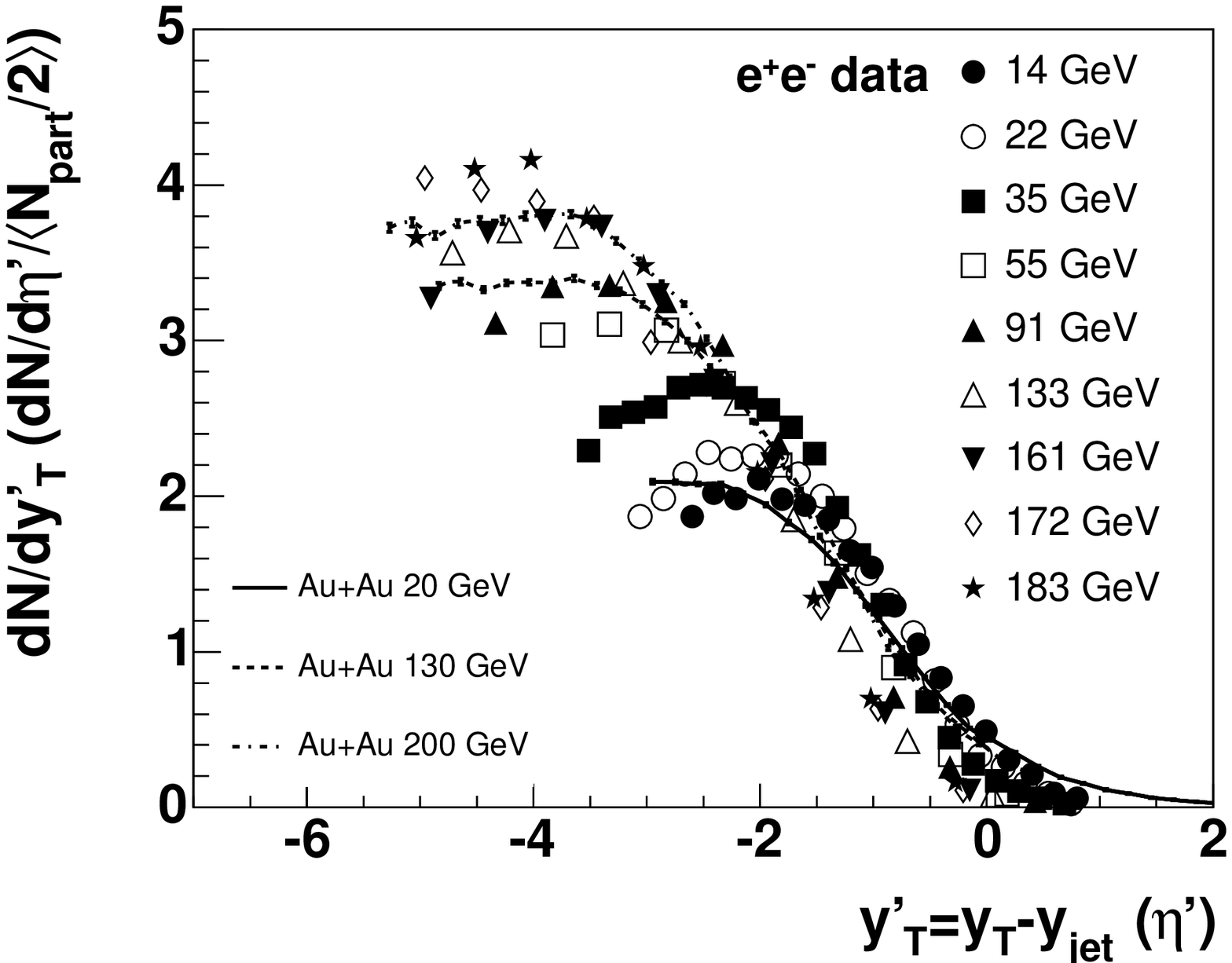}
\caption{
$\dndetapnp$ for $\AAA$ collisions compared to $dN/dy^{\prime}_T$
in $\epem$ reactions over a broad range of $\s$.
\label{ee_AA_limfrag_bw}
}
\end{center}
\vspace*{-1cm}
\end{figure}

The previous discussions have suggested that rapidity distributions
are not merely the consequence of the nucleon or nuclear wave functions,
but may be dynamically generated by Landau initial conditions.
The overall similarity between the pseudorapidity distributions in $\AAA$
and $\epem$ collisions at the same $\s=200$ GeV (shown in Fig.~\ref{ee_AA_limfrag_bw} in the limiting fragmentation frame) 
certainly complicate
separate explanations of these systems.  It is also observed that
the particle density per participant pair in $\AAA$ is similar
to the density relative to the thrust axis in $\epem\rightarrow
hadrons$.  Given these similarities, the observation of limiting
fragmentation in $\pp$ and $\epem$ (also shown in Fig.~\ref{ee_AA_limfrag_bw},
although less precise than that observed in $\AAA$) 
is not surprising.

\begin{figure}[t]
\begin{minipage}{75mm}
\includegraphics[width=75mm]{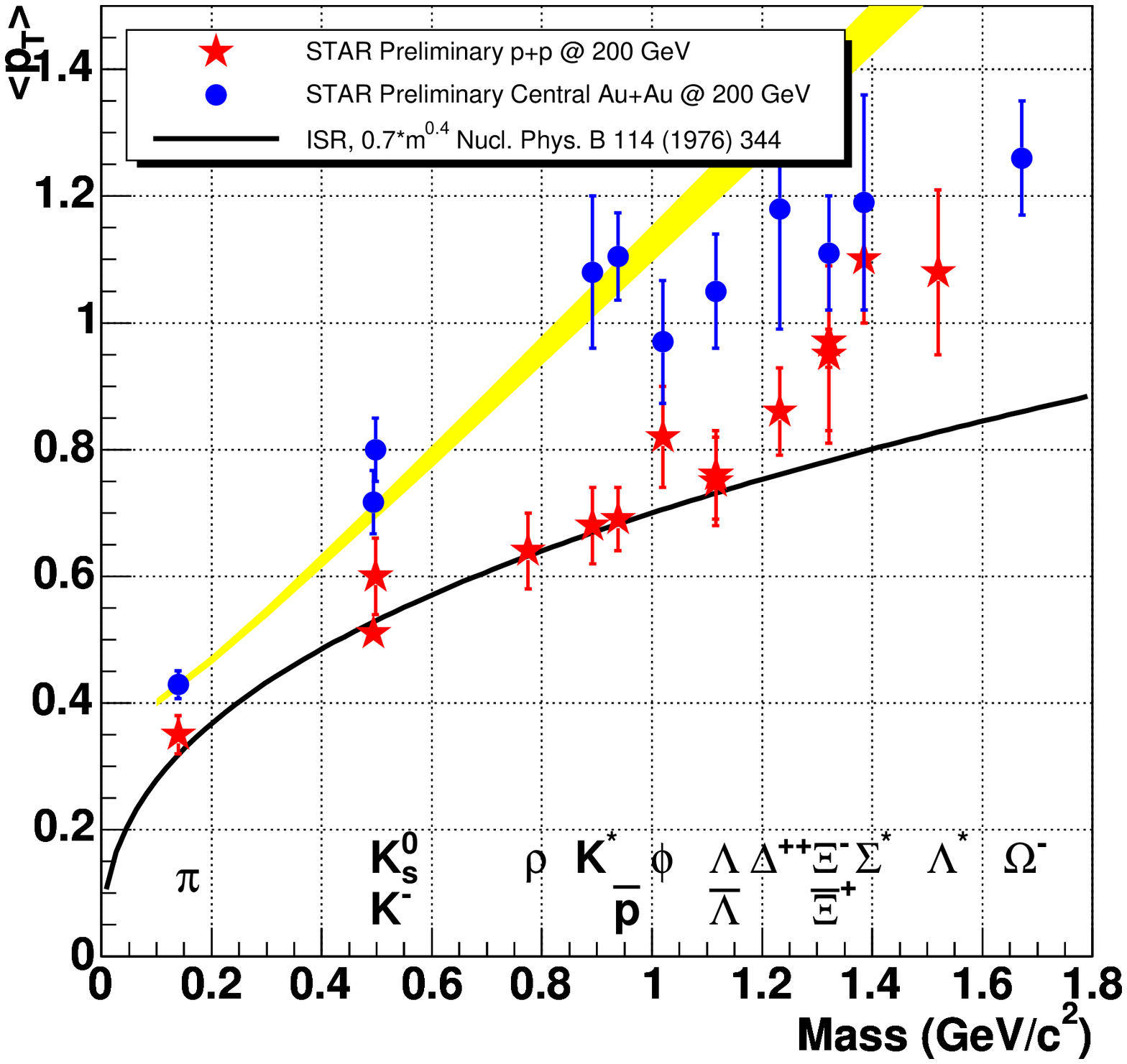}
\caption{
Preliminary data on
$\mpt$ measured in $\AAA$ and $\pp$ collisions at $\s=200$ GeV.
\label{star_meanpt_vs_mass}}
\end{minipage}
\hspace{\fill}
\begin{minipage}{75mm}
\includegraphics[width=75mm]{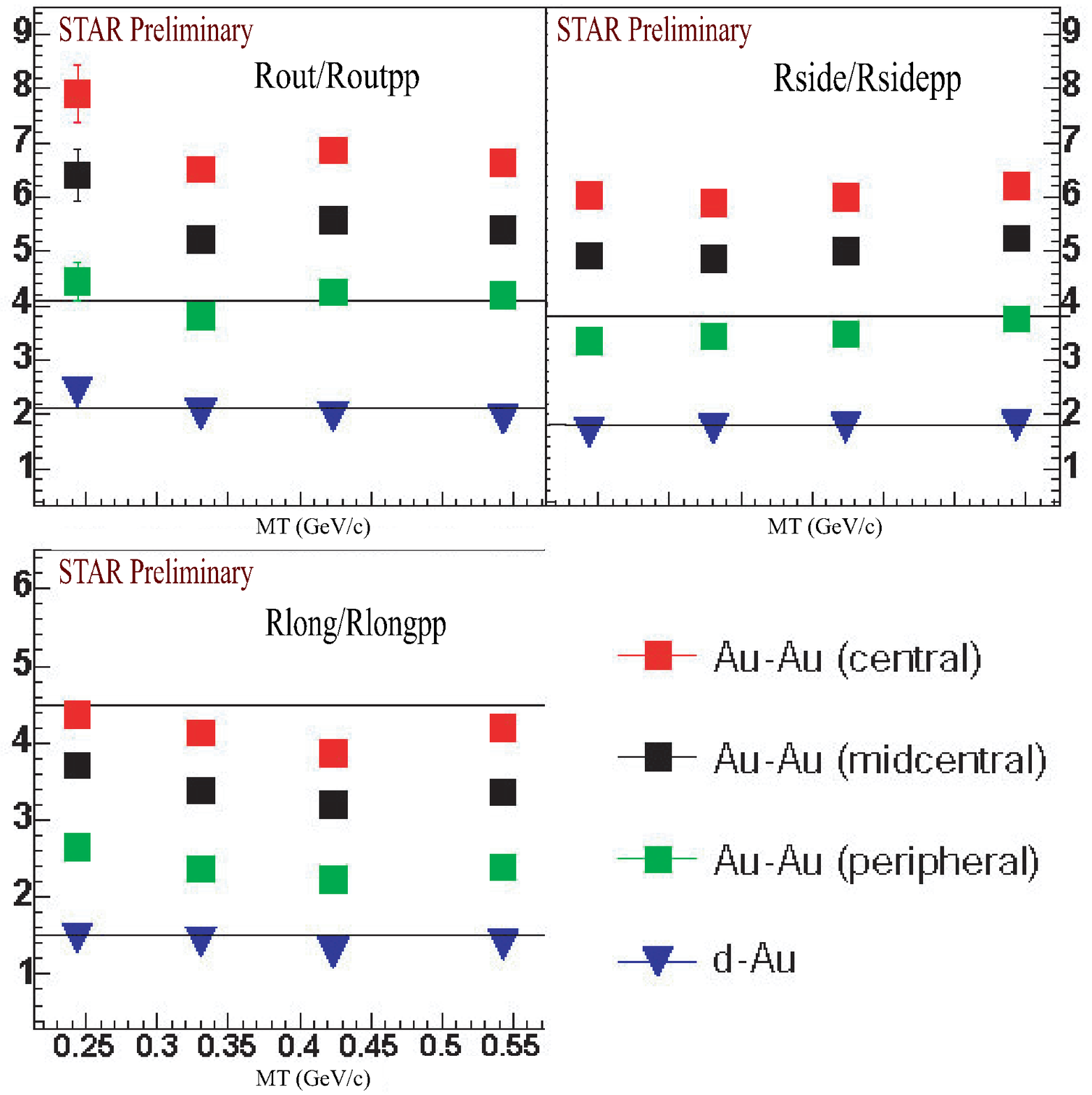}
\caption{
HBT radii as a function of $k_T$ for d+Au and Au+Au collisions
divided by the radii measured in $\pp$ collisions.
\label{dAuAuAuratiomt_2}
}
\end{minipage}
\vspace*{-.7cm}
\end{figure}

However, the issue of whether or not these systems are truly
equilibrated is contingent on whether or not collective behavior
can be discerned.  It is well-known that the particle
yields in $\epem$ and $\pp$ collisions are described by statistical models
at the same chemical freezeout temperature as $\AAA$ at RHIC
energies~\cite{Cleymans:2002mp,Braun-Munzinger:2003zz}.
However, it is a subject of debate whether or not this
is simply ``phase space dominance'' as opposed to a true equilibration
process~\cite{Koch:2002uq}. 
Thus, the small systems are not regarded as truly collective.
And yet, recent preliminary STAR data on identified spectra in 
$\pp$ collisions, shown in Fig.~\ref{star_meanpt_vs_mass}~\cite{star-mt},
shows a qualitatively similar rise in the mean
$p_T$ as a function of particle mass.  Other preliminary data on
HBT correlations, shown in Fig.~\ref{dAuAuAuratiomt_2}~\cite{star-hbt}, 
finds that the pronounced 
$k_T$ dependence of HBT radii in Au+Au, 
which is normally understood to yield information on radial expansion, 
is proportional to that found in d+Au and $\pp$.  Thus, we are
faced the situation that experimental observables are interpreted as 
expansion dynamics in the case of $\AAA$, but as merely reference data
in the more elementary systems.  


\section{Conclusions}

It is argued in this proceedings that soft physics in strongly
interacting systems may well be simpler than typical dynamical models
would generally suggest.  In fact, hydrodynamics may provide a 
unified conceptual framework from the first moment of the collision
all the way to freezeout.  More than that, it may provide a basis
for understanding elementary systems as well, perhaps complementing the
approach of understanding these systems with perturbative QCD.
Of course, QCD will be fundamental in understanding the basic
degrees of freedom of these strongly-interacting systems and how
they thermalize so quickly, as implied by the Landau approach.
And yet, basic issues purely related to hydrodynamics, such as
the integration of longitudinal and transverse dynamics will
have to be tackled for real progress to be made.  Finally, 
while the focus of RHIC physics has been at mid-rapidity, 
detailed understanding of the global features of the event 
across the whole rapidity range will be necessary to fully
understand the collective nature of these collisions.

\section{Acknowledgments}
The author would like to thank the organizers of the heavy ion session
at INPC2004, Hans-Ake Gustafsson and Jens J. Gaardhoje, for the 
invitation to speak at the conference.  Additional thanks to Richard Witt for providing Fig.~\ref{star_meanpt_vs_mass}, and Mark Baker for useful discussions.

\end{document}